\documentclass[prl,nofootinbib,reprint,superscriptaddress]{revtex4-2}

\UseRawInputEncoding

\usepackage{amsmath}
\usepackage{amsfonts}
\usepackage{color}
\usepackage{comment}
\usepackage{diagbox}
\usepackage{graphicx}
\usepackage{wasysym} 
\usepackage{xspace}

\graphicspath{{../fig/}{./}}

\newcommand{\ie}{{i.e.,}\xspace}
\newcommand{\eg}{{e.g.}\xspace}

\newcommand{\acronym}[1]{{\text{\uppercase #1}}\xspace}
\renewcommand{\acronym}[1]{{\text{\MakeUppercase{#1}}}\xspace}
\newcommand{\rabbit}{\acronym{rabbit}}

\newcommand{\mfrog}{mixed-\acronym{FROG}}
\newcommand{\mFROG}{\mfrog}
\newcommand{\kraken}{\acronym{KRAKEN}}

\newcommand{\tdse}{\acronym{tdse}}

\newcommand{\xuv}{\acronym{xuv}}
\newcommand{\XUV}{\xuv}

\newcommand{\vmax}{\ensuremath{v_{\mbox{\tiny max}}}\xspace}

\newcommand{\rdm}{\acronym{rdm}}
\newcommand{\RDM}{\rdm}
\newcommand{\ket}[1]{\ensuremath{\vert #1 \rangle}\xspace}

\newcommand{\tauxx}{\ensuremath{\tau_{\text{\small xuv}}}\xspace}

\newcommand{\Ene}{\ensuremath{\varepsilon}\xspace}

\newcommand{\Ytot}{\ensuremath{Y_{\text{\tiny tot}}}\xspace}

\newcommand{\psitot}{\ensuremath{\ket{\psi_{\text{\tiny end}}}}\xspace}
\newcommand{\rhotot}{\ensuremath{\rho_{\text{\tiny end}}}\xspace}
\newcommand{\ptot}{\ensuremath{p_{\text{\tiny end}}}\xspace}
\newcommand{\phiE}{\ensuremath{\ket{\varphi_\Ene}}\xspace}
\newcommand{\chiv}{\ensuremath{\ket{\chi_v}}\xspace}
\newcommand{\chim}{\ensuremath{\ket{\tilde\chi_m}}\xspace}

\newcommand{\trace}[1]{\ensuremath{\mbox{tr}(#1)}\xspace}

\definecolor{jccol}{rgb}{1,0,0}
\definecolor{mbcol}{rgb}{0.75,0.0,0.75}
\definecolor{rtcol}{rgb}{0,1,0}
\definecolor{clcol}{rgb}{0,0,1}

\definecolor{done}{rgb}{0,0.8,0}
\definecolor{tobedone}{rgb}{0.8,0,0}

\newcommand{\LCPMR}{Sorbonne Universit\'e, CNRS, Laboratoire de Chimie Physique-Mati\`ere et Rayonnement, LCPMR, 75005 Paris, France}
\newcommand{\Attolab}{\modif{Universit\'e Paris-Saclay, CEA, LIDYL}, 91191 Gif-sur-Yvette, France}
\newcommand{\IOGS}{Laboratoire Charles Fabry, Institut d'Optique Graduate School, CNRS, Universit\'e Paris-Saclay, 91127 Palaiseau, France}

\newlength{\figwidth}

\newcommand{\modif}[1]{\textcolor{jccol}{#1}}
\renewcommand{\modif}[1]{#1}

\begin{document}
\setlength{\figwidth}{0.99\linewidth}

\title{
Complete retrieval of attosecond photoelectron dynamics from  partially-coherent states in entangled photoemission
}

\author{Morgan Berkane}
\affiliation{\LCPMR}
\author{Richard Ta\"\i eb}
\affiliation{\LCPMR}
\author{Gabriel Granveau}
\affiliation{\IOGS}
\author{Pascal Sali\`eres}
\affiliation{\Attolab}
\author{Charles Bourassin-Bouchet}
\affiliation{\IOGS}
\author{Camille L\'ev\^eque}
\affiliation{\LCPMR}
\author{J\'er\'emie Caillat}
\email{jeremie.caillat@sorbonne-universite.fr}
\affiliation{\LCPMR}

\date{\today}

\begin{abstract}
We show that the complete photoemission dynamics in situations of electron-ion entanglement can be retrieved from photoelectron spectral measurements without information on the ion.
To this end, we develop an energy-time analysis of the photoelectron's reduced density matrix based on first principles.
We test and assess our approach with numerical simulations on a low dimensional model molecule in interaction with broadband composite pulses occulting the vibrational resolution. Our method is directly applicable to recent experimental schemes measuring the photoelectron reduced density matrices in atomic and molecular photoemission. Therefore, it opens a new window on the dynamics of decoherence and entanglement at the attosecond timescale.
\end{abstract}
\maketitle

Attosecond time-resolved spectroscopies rely to a large extent on inferferometric schemes~\cite{paul2001a,hentschel2001a,krausz2009a}, where optical and `quantum' phases~\cite{veniard1996a,muller2002a,haessler2009a,schultze2010a,yakovlev2010a,klunder2011a} play a central role in the measurements and their interpretations. Standard approaches, such as the \rabbit~\cite{paul2001a} and the attosecond streaking~\cite{hentschel2001a}, as well as their expanding lineage~\cite{chini2010a,gruson2016a,borrego-varillas2022a}  assume totally coherent and fully resolved processes and measurements for the time-domain interpretation of experimentally measurable spectral amplitudes. However, the question of decoherence is ubiquitous in this context. It has multiple origins ranging from technical causes to quantum entanglement and statistical mixtures, in the preparation of the studied systems~\cite{rosca-pruna2001a,boutu2008a,medisauskas2015a,li2008a,monfared2022a},  the measurements~\cite{bourassin-bouchet2020a} or the investigated processes themselves~\cite{goulielmakis2010a,pabst2011a,vacher2017a,jordan2020a,vrakking2021a,cattaneo2022a,koll2022a,vrakking2022a,nabekawa2023a,blavier2022a,maxwell2022a,ruberti2022a,nandi2024a}. They have long been addressed only through indirect signatures in the observations. Recently, new pump-probe approaches have been designed to explicitly account for various potential sources of decoherences, in particular the \mFROG~\cite{bourassin-bouchet2015a,bourassin-bouchet2020a} and \kraken~\cite{laurell2022a,laurell2023a}. They give access to the photoelectron's {\em final} reduced density matrix (\rdm). The problem of recovering the comprehensive photoemission dynamics from an asymptotic \rdm still remains  unsolved. \modif{This would give access to the ultrafast build-up of entanglement and decoherence in real time.}

In this Letter, we demonstrate the possibility to reconstruct unambiguously the entangled photoemission dynamics out of a photoelectron's \rdm. 
 Our reasoning borrows from quantum information the concept of mixed quantum state purification~\cite{schrodinger1936a,hughston1993a}  and from attosecond science an energy-time analysis of interferometric measurements in the spectral domain~\cite{desrier2018a}. We illustrate our approach with numerical simulations inspired by an original physical case first proposed  in~\cite{vrakking2021a} and further explored in~\cite{koll2022a,vrakking2022a,nabekawa2023a}. In this scheme, a diatomic molecule is ionized  by a pair of \XUV pulses whose temporal separation \tauxx serves as a knob to tune the final photoelectron-photoion entanglement through its (mis)match with the molecule's vibrational eigenperiods. With our method, we evidence an ultrafast build-up of entanglement-induced decoherence during light-matter interaction.

Our model  is a two-dimensional molecule~\cite{caillat2018a}, denoted A$_2$ in the following, with one degree of freedom accounting for the electron motion, and another one  for the molecular vibration. It is therefore, by construction, an idealized ``photoelectron+photoion'' bi-partite system. It has been previously used as a numerical benchmark in various studies, see\modif{, \eg,} Refs.~\cite{houfek2006a,caillat2011a,labeye2024a,desrier2024a}. Here we use the same parameters as in~\cite{labeye2024a} which mimic, to some extent, the properties of H$_2$ -- namely the vibronic energies in its ground electronic state and first ionic state. 

Owing to the symmetry of the molecule along the electron's coordinate, and thanks to the dipole selection rules in one dimension, the photoelectron state can be uniquely identified by its energy $\Ene\in\mathbb{R}^+$ (\ie the photoemission direction is here discarded). Regarding the nuclear motion, we characterize the ion's state by its vibrational level $v\in\mathbb{N}$. Hence, the present work is focused on analyzing the outcome of the molecular photoemission process 
\begin{eqnarray}
\mbox{A}_2+\hbar\omega(\tauxx) &\rightarrow&\mbox{A}_2^+(v) + \mbox{e}^-(\Ene).
\end{eqnarray}
We set the ionizing \xuv central frequency to $\omega=54.29$~eV (35\textsuperscript{th} harmonic of a 800-nm laser). We assigned each of the two pulses a short $\sin^2$ envelop lasting $1.60$~fs (full duration), in the perturbative intensity regime. We simulated the process by solving  the time-dependent Schr\"odinger equation (\tdse) with a numerically exact scheme~\cite{caillat2018a} consisting in expanding the vibronic wave-function on the orthonormal basis of ionic vibrational states, $\{\chiv\}$, and propagating the associated channel-dependent electron wave-packets. Here, we obtained converged results by including all vibrational channels up to $\vmax=16$. Using the  electronic continuum manifold basis $\{\phiE\}$, the final state of the ``photoelectron+photoion'' system can be expressed as\footnote{ Equations are expressed in atomic units (a.u.) all through the Letter. }
\begin{eqnarray}\label{eqn:psitot}
\psitot&=&\int\limits_0^{\infty}\,\sum\limits_{v=0}^{\vmax}A_v(\Ene)\,\phiE\otimes\chiv\,d\Ene.
\end{eqnarray}
We fully characterized the photoemission process in our simulations by computing the channel-dependent spectral amplitudes $A_v(\Ene)$ out of the continuum wave-packets
~\cite{caillat2005a,desrier2024a}, in the asymtotic region. 
\begin{figure}[t]
\center
\includegraphics[width=\linewidth]{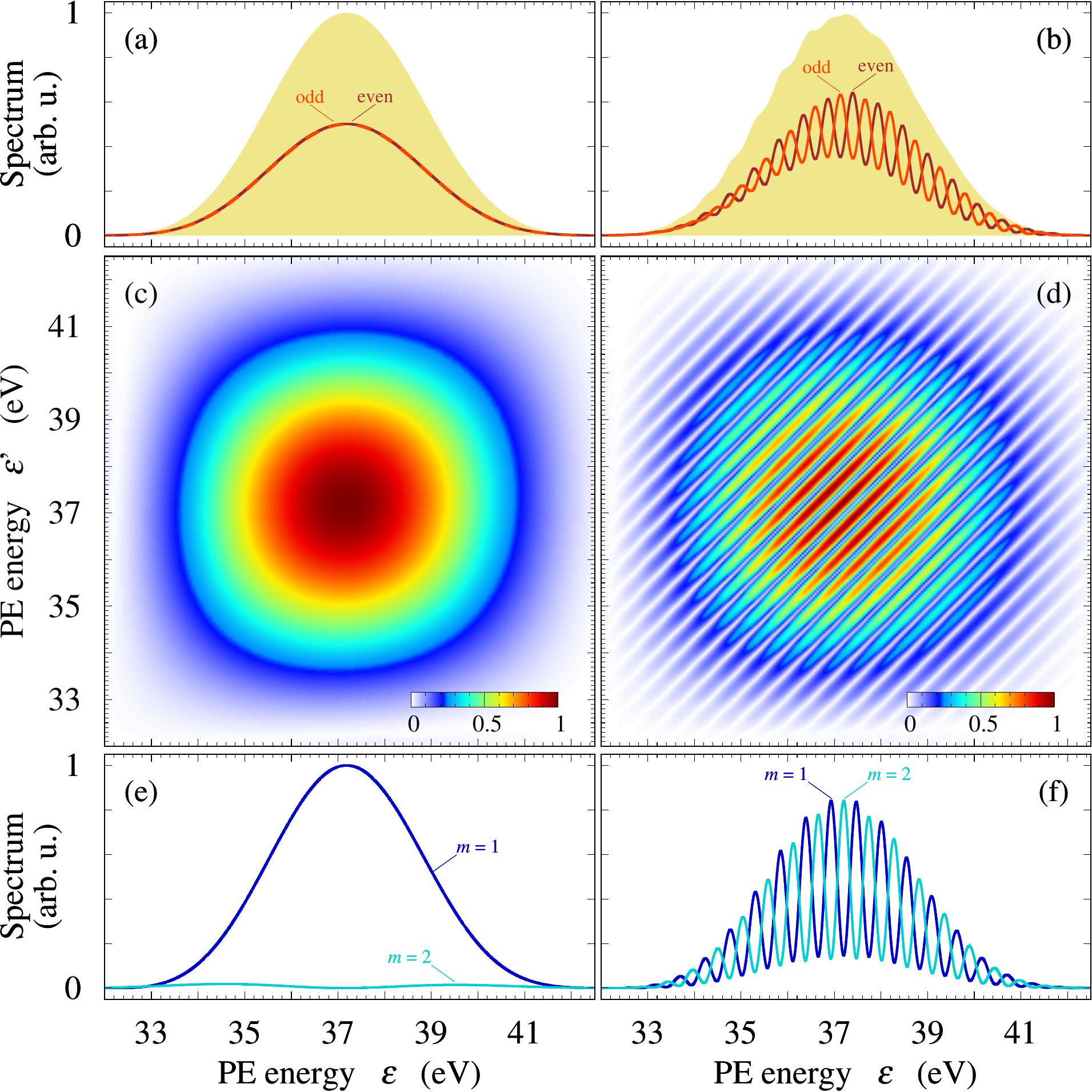}
\caption{\label{fig:spectral} Photoelectron characterization in the spectral domain, in the $\tauxx=0$ (case I, left column) and $\tauxx=\frac{1}{2}T_{01}$  (case II, right column) cases, respectively. (a),(b): Integrated PE spectra $\sigma(\Ene)$ (yellow filled curve), and its even- and odd-$v$ components (see labels); (c),(d): Modulus of the photoelectron's \RDM, $\vert \rho(\Ene,\Ene')\vert $; (e),(f):  Spectra of the main Schmidt modes, $s_1(\Ene)$ and $s_2(\Ene)$ (see label). The data were normalized to 1 at the maximum of $\sigma(\Ene)$ in the $\tauxx=0$ case. 
}
\end{figure}

To illustrate the interplay between \tauxx and the final entanglement of the ionized system~\cite{vrakking2021a}, \modif{we simulated} two characteristic values of \tauxx: 0 fs (case I) and $\frac{1}{2}T_{01}\simeq 7.62$ fs (case II). The latter is half the vibrational period $T_{01}\equiv2\pi/\Delta\mathcal{E}_{01}$ associated with the energy gap $\Delta\mathcal{E}_{01}=271$~meV between the first two channels. For the sake of comparison, we multiplied the field strength by $\sqrt{2}$ in case II to obtain the same final ionization yields as in case I.  While previous studies~\cite{vrakking2021a,koll2022a,nabekawa2023a} addressed partial measurements based on {\em  photoion} spectroscopy,  we focus here on the complementary case where only the {\em photoelectron} (PE) is detected and characterized through its final \rdm. 

\begin{table}[t]
\begin{tabular}{cccccc}
\hline
$m$ & 1 & 2 & 3 & 4 & $\geq 5$ \\ \hline
(a) & $97.6\%$ & $ \textcolor{white}{0}2.3\%$ & $<0.1\%$ & $<0.1\%$ & $<10^{-3}\%$ \\ 
(b) & $49.4\%$ & $47.9\%$ & $\textcolor{white}{0}1.4\%$& $\textcolor{white}{0}1.1\%$ & $\lesssim0.1\%$ \\ \hline
\end{tabular}
\caption{\label{tab:schmidt} Weights $W_m$ of the photoelectron Schmidt modes in the (a) $\tauxx=0$ and (b) $\tauxx=\frac{1}{2}T_{01}$ cases. They are expressed in percentages of the final ionization yields $\Ytot$.}
\end{table}

\modif{We  first consider the $v$-integrated photoelectron spectra (PES) computed as
\begin{eqnarray}\label{eqn:PES}
\sigma(\Ene)&=&\sum\limits_{v=0}^{\vmax}\vert A_v(\Ene)\vert^2.
\end{eqnarray}}
Fig\modif{ure}~\ref{fig:spectral}(a) shows the spectrum (yellow filled curve) obtained for case I.
It displays a single broad peak whose shape is dominated by the pulse bandwidth (approximatively 5~eV at half maximum) which significantly exceeds the spectral separation of the vibrational channels (few $0.1$ eV). 
The spectrum obtained for case II is shown in \modif{Fig.~\ref{fig:spectral}}(b). It presents the same globally featureless bell shape, with only weak oscillations. Hence, the PE spectra alone reveal no clear signature of the \tauxx-dependent photoelectron-photoion entanglement.

The difference is striking when looking at the photoelectron's asymptotic \rdm, 
\begin{eqnarray}\label{eqn:RDM}
\rhotot(\Ene,\Ene')&=&\sum\limits_{v=0}^{\vmax} [A_v(\Ene)]^\star A_v(\Ene'),
\end{eqnarray}
where $[\ ]^\star$ denotes the complex conjugate. It is implicitly expressed on the photoelectron basis $\{\phiE\}$, and its diagonal $\rhotot(\Ene,\Ene)$ corresponds to the spectrum $\sigma(\Ene)$, Eq.~\eqref{eqn:PES}. Relevant additional information is provided by the off-diagonal features, namely the coherences between the populated energies. The modulus of the \rdm obtained in case I, shown in Fig.~\ref{fig:spectral}(c), is quasi-circular. Notably, it displays a similar structuration in the $\Ene=\Ene'$ and the $\Ene=-\Ene'$ directions. This is a direct signature of a weakly entangled state, as could be expected in molecular photoemission with a single broadband pulse: detecting a photoelectron at a given energy \Ene  provides no information regarding the state $v$ of the ion.

In contrast, the \rdm obtained in case II displays a clear banded structure, see \modif{Fig.~\ref{fig:spectral}}(d). Here, the structuration radically differs in the $\Ene=\Ene'$ and the $\Ene=-\Ene'$ directions, indicating a strong entanglement. Indeed, a non-vanishing delay between the two  identical pulses induces interferences in the process, with a channel-dependent character~\cite{vrakking2021a}.  
  With this choice for \tauxx, the spectral alternance of constructive and destructive interferences in the odd and even parity channels are nearly out of phase, leading to observed oscillations along the antidiagonal of the \rdm.  This can be verified by looking at the channel-parity-resolved spectra, which can be isolated {\em in our simulations}, see  Fig.~\ref{fig:spectral}(b) (solid lines, see labels). These spectra appear as two interspersed combs, each of them with a $2\pi/\tauxx$ period. The phase opposition would be exact all through the spectral range if the ionic potential were  harmonic. In contrast, the parity-resolved components of the spectrum for case I are practically undistinguishable and as featureless as the integrated spectrum itself. To summarize, case II creates a significantly entangled photoelectron-photoion state, with nevertheless little signatures of the entanglement in the sole photoelectron spectrum measured {\em without vibrational resolution}. 

The decoherence induced by this entanglement can be further quantified through the purity $\ptot=\trace{\rhotot^2}/\trace{\rhotot}^2$, where $\trace$ denotes the trace operation~\cite{tichy2011a}. It is nearly maximum in case I ($\ptot=0.95$), while it has approximately half that value in the other case ($\ptot=0.48$). 

\begin{figure*}[t]
\center
\includegraphics[width=0.9\linewidth]{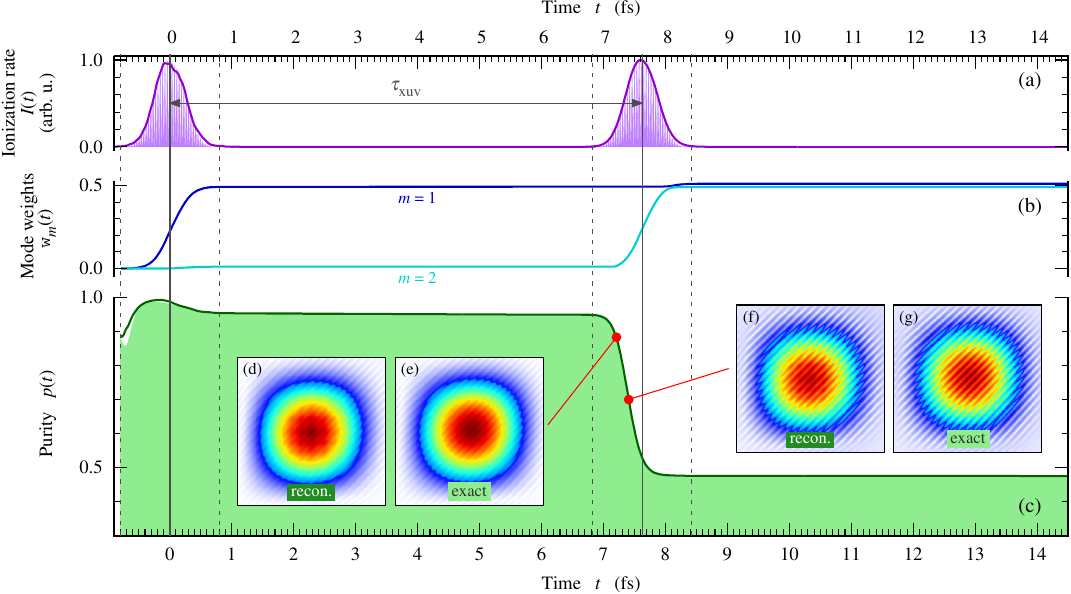}
\caption{\label{fig:temporal} Photoelectron characterization in the time domain in the $\tauxx=\frac{1}{2}T_{01}$ case . (a): Reconstructed ionization rate $I(t)$ (dark violet curve) and instantaneous \xuv vector potential squared (violet filled curve). (b): Reconstructed weights $w_m(t)$ for the  two dominant modes (see labels). (c):  Reconstructed time-dependent purity $p(t)$ (dark green curve) and numerically exact purity (light green filled curve). (d) and (e): Modulus of the reconstructed (d) and exact (e) photoelectron's \rdm $\vert\rho(\Ene,\Ene';t)\vert$ at $t=7.21$ fs.  (f) and (g): Same at $t=7.41$ fs. The {\rdm}s are displayed over the same spectral range as in Fig.~\ref{fig:spectral}(c) and (d). For each pulse, the central time is indicated by a vertical solid line and the beginning and end times by dashed vertical lines. The time origin $t=0$ is set to the maximum of the first pulse.}
\end{figure*}

To proceed with the analysis of the \rdm, we use the so-called purification,  a general procedure applying to mixed quantum states~\cite{schrodinger1936a,hughston1993a}. It consists here in considering the Schmidt modes of the photoelectron, \ie the normalized eigenvectors $f_m(\varepsilon)$ of the \RDM with non-vanishing eigenvalues ${\Lambda_m}\in\mathbb{R}^+$~\cite{horodecki2008a,blavier2022a}. They allow for a low-rank reformulation of \psitot [Eq.~\eqref{eqn:psitot}] as
\begin{eqnarray}\label{eqn:psipur}
\psitot&=&\int\limits_0^{\infty}\,\sum\limits_{m=1}^{M} \underbrace{\sqrt{\Lambda_m}f_m(\Ene)}_{=: B_m(\Ene)}\phiE\otimes\chim\,d\Ene
\end{eqnarray}
with $M\leq\vmax+1$, and where 
$\{\chim\}$ is an effective, {\em a priori} unspecified, orthonormal vibrational basis for the final state. Note that this formal expression is very general and holds as well when the {\em nature} of the unresolved degree(s) of freedom is unknown. This  encompasses a broad range of practical situations. 
 In contrast to the unresolved amplitudes $A_v(\Ene)$, the modes $B_m(\Ene)$ are orthogonal by construction, \ie  $\int  [B_m(\Ene)]^\star B_{m'}(\Ene)d\Ene=\delta_{mm'}\Lambda_m$.  
 We give in Tab.~\ref{tab:schmidt} the leading values of $\Lambda_m$ obtained in the two cases, in percentages of the total ionization yields $\Ytot=\trace{\rhotot}=\sum_{m=1}^{M}\Lambda_m$. The  spectra
\begin{eqnarray}\label{eqn:sigm}
s_m(\Ene)&=&\vert B_m(\Ene) \vert^2
\end{eqnarray}
 associated with the first two modes are shown in the lower frames of Fig.~\ref{fig:spectral}. The number of effective modes in each case is consistent with the final purities commented above, having in mind that the lower bound for the latter is given by $1/M$. Indeed, in case I, the $m=1$ mode largely dominates, by more than $97\%$.  As seen in Fig.~\ref{fig:spectral}(e), the bell-shaped dominant mode clearly accounts for the global trend of the  integrated spectrum $\sigma(\Ene)$ shown in Fig.~\ref{fig:spectral}(a). Such a nearly rank-one \RDM is an alternative signature of a nearly separable, \ie non-entangled, state.  In case II, the $m=1$ and $2$ modes are dominating with comparable magnitudes. Their spectra $s_m(\Ene)$, shown in Fig.~\ref{fig:spectral}(f), are shaped as two shifted combs reminiscent of the channel-parity-resolved spectra~\cite{koll2023a} with significantly more contrasted oscillations, see \modif{Fig.~\ref{fig:spectral}}(b).  These observable $m$-selective spectra therefore fully reveal the entangled physics induced, in the unresolved degree of freedom, by the \xuv pulse configuration which approximately discriminates the $v$ parity. 

Beyond the individual physical interpretations of the Schmidt modes, their spectra sum up to the diagonal of the \rdm,  $\sigma(\Ene)$. These $B_m(\Ene)$ modes are directly accessible from the measurement of \rhotot even in partially resolved experiments. Furthermore, they are complex-valued  amplitudes that can be analyzed in the time-domain, by generalizing tools originally designed for attosecond time-resolved interferometric schemes assuming fully coherent processes~\cite{busto2018a,desrier2018a}. From there on, we hence focus on the spectro-temporal analysis and interpretations of the Schmidt modes $B_m(\Ene)$. We will illustrate this with case II, and validate the ``reconstructed'' dynamics by confronting them to the actual dynamics extracted from the numerical solution of the \tdse. 

To this end, we introduce the temporal amplitudes~\footnote{We use here the same phase and normalization conventions as in~\cite{desrier2018a}.}
\begin{eqnarray}\label{eqn:bmt}
b_m(t)&:=&\int\limits_{-\infty}^{+\infty}\,B_m(\Ene)\,e^{-i\Ene t}\,d\Ene. 
\end{eqnarray}
Note that, to highlight the Fourier nature of the analysis, we shifted  the lower integration bound from $0$ (in Eq.~\ref{eqn:psitot}) to  $-\infty$,
the support of $B_m(\Ene)$ being restricted to the positive energies. We then build spectro-temporal amplitudes out of $b_m(t)$ as
\begin{eqnarray}
\mathcal{B}_m(\Ene;t)&:=&\frac{1}{2\pi}\int\limits_{-\infty}^{t}\,b_m(t')\,e^{+i\Ene t'}\,dt'.
\end{eqnarray}
The amplitudes $b_m(t)$ and $\mathcal{B}_m(\Ene,t)$ are mathematical analogues of the quantities derived out of asymptotic spectral amplitudes in single-channel~\cite{gruson2016a} or fully resolved~\cite{autuori2022a} experimental investigation of atomic photoemission dynamics. Among the various time-energy analysis invoked in similar contexts (see\modif{, \eg,}~\cite{busto2018a}), the considered one presents the advantage of rigorously giving access to fundamental physical quantities, such as ionization rates and transient photoelectron spectra, based on first principles of quantum physics. Hereafter, we generalize these interpretations to the present context of partially resolved photoemission. They were shown in~\cite{desrier2018a} to be valid for photoemission processes towards continua with negligible spectral variations, \ie not too close to the threshold. These conditions are met in our simulations.

We begin with the ionization rate which, in the present context, can be evaluated as
 \begin{eqnarray}\label{eqn:recrate}
I(t)&=&\frac{1}{2\pi}\sum_{m=1}^{M}\vert b_m(t)\vert^2.
 \end{eqnarray}
 Here and all through the rest of the paper, we included all the modes (\ie $M=\vmax+1$) in the temporal reconstructions for the sake of numerical accuracy. The time-dependent rate reconstructed as such in case II is shown as a  dark-violet curve in Fig.~\ref{fig:temporal}\modif{(a)}. It appears as a succession of two bell-shaped structures coinciding with the pulse profile, also shown as a filled violet curve. Rigorously, the reconstructed profile is time-shifted by the Wigner delay associated with the photoelectron scattering in the continuum~\cite{pazourek2015a,desrier2018a}. However, in the considered 1-photon process towards a smooth continuum far from threshold, this delay is of few attoseconds and cannot be resolved in the figure. The excellent agreement between the reconstructed rate and the pulse profile is therefore a first validation of our time-domain reconstruction scheme. 

As for the $\mathcal{B}_m(\Ene;t)$ amplitudes, they collectively give access to the temporal build-up of the photoelectron's \rdm itself,  reconstructed  at each time $t$ as
\begin{eqnarray}\label{eqn:exaRDM}
\rho(\Ene,\Ene';t)&=&\sum\limits_{m=1}^{M} [\mathcal{B}_m(\Ene;t)]^\star \mathcal{B}_m(\Ene';t).
\end{eqnarray} 
  It is important to note that the physical meaning of the individual $\mathcal{B}_m(\Ene;t)$  is limited. In particular they should not be confused with the time-dependent \rdm's Schmidt modes. We  obtained the latter, and the corresponding eigenvalues ${\lambda_m(t)}$, by diagonalizing the reconstructed $\rho(\Ene,\Ene';t)$ at each time. We show in Fig.~\ref{fig:temporal}(b) the normalized weights  $w_m(t)=\lambda_m(t)/\Ytot$. Since no significant field-free, vibronically correlated dynamics is likely to take place in this model molecule, they display visible variations only in presence of the \xuv pulses.  
The values of $w_m(t)$ reached between the two pulses are consistent with the final -- properly rescaled -- weights ($W_m$) obtained in the single pulse configuration (see Tab.~\ref{tab:schmidt}).  Out of the reconstructed \rdm, we also computed the time-dependent purity, $p(t)=\trace{\rho(t)^2}/\trace{\rho(t)}^2$, shown as a dark-green curve in Fig.~\ref{fig:temporal}(c). As for the Schmidt mode weights, the purity variations are strictly concomitant with the interactions with the light pulses. Starting slightly below $0.9$, $p(t)$ keeps a relatively high value during the first pulse and ends up at $0.95$, \ie the value of \ptot obtained in case I. Consistently with the scheme design, an important purity \modif{variation} occurs with\modif{in} the second pulse, where it decays monotonically towards its final value of $0.48$. This ultrafast decoherence dynamics is characterized by a maximum entanglement speed of $-dp/dt=1.07$ fs\textsuperscript{-1} at $t=7.40$~fs. Interestingly, $p(t)$ evolves mostly in the {\em first half} of  each pulse, as it depends quadratically on the mode weights, while the modes themselves continue to evolve \modif{significantly} until the end of the pulse.

To assess the accuracy of the reconstructed $p(t)$, we considered the numerically exact time-dependent {\RDM} of the photoelectron directly obtained in the \tdse simulations. We computed it by analyzing the outgoing photoelectron wave-packets using absorbing boundaries~\cite{caillat2005a}, shortly after the end of both pulses to avoid any spurious field dressing artifact while keeping negligible any temporal distorsion due to the wave-packet spreading. The corresponding purity is shown in the same figure as a light-green filled curve. 
 The agreement with the reconstructed $p(t)$ is remarkable all through the simulated time range, with minor discrepancies showing up only at the earliest times when the ionization probability has reached less than $1\%$ of the final yields.  We verified that the agreement holds not only for the purity but also for the whole time-dependent \RDM itself. It evolves from the unstructured quasi-circular shape shown in Fig.~\ref{fig:spectral}(c) (before the arrival of the second pulse) to the striped shape shown in Fig.~\ref{fig:spectral}(d) (after the second pulse). This is illustrated at two arbitrary times during the second pulses in  Fig.~\ref{fig:temporal}(d-g), revealing the progressive build-up of the diagonal stripes. The excellent agreement between the reconstructed and the actual \RDM in these two cases is  representative of the quality of the procedure at all times. This is a numerical demonstration of the capacities of our time-energy analysis to provide a complete picture of the actual ionization dynamics from the photoelectron perspective, at the natural, intra-pulse, attosecond time-scale of the process. 
 
To conclude, we have introduced an original method to reconstruct complete photoemission dynamics  out of the photoelectron's asymptotic reduced density matrix in the spectral domain. It is based on an {\em ad hoc} `time-frequency' analysis of the photoelectron \rdm's Schmidt modes.  We illustrated it with numerical simulations on a  model molecule ionized by composite \xuv pulses. An excellent agreement is obtained between the \RDM-based reconstructed dynamics and the actual ones that are fully accessible in the simulations. 
From the experimental point of view, the present method requires the measurement of the photoelectron \RDM which is now possible with recently developed approaches such as the \mFROG~\cite{bourassin-bouchet2020a} or  \kraken~\cite{laurell2022a}. 
It can thus be applied to a variety of photoemission processes -- that play a key role in attosecond time resolved spectroscopies -- regardless of the nature of the unresolved degree(s) of freedom (\modif{\eg,} rotational, vibrational or electronic).  
It opens the perspective of following experimentally, in ``real time'', the ultrafast buildup of entanglement-induced decoherence due to interchannel couplings~\cite{pabst2011a,medisauskas2015a}, with important consequences for\modif{, \eg,} charge migration and attochemistry~\cite{cardosa-gutierrez2024}.

\begin{acknowledgments}
This research received the financial support of the French National Research Agency through Grant No. ANR-20-CE30-0007-DECAP.
\end{acknowledgments}

\bibliography{../../biblio/jcbib}

\end{document}